\documentclass[twocolumn]{aastex631}

\received{March 8, 2024}
\revised{June 5, 2024}
\accepted{June 24, 2024}

\acceptedjournal{Astronomical Journal}

\shorttitle{Two Exoplanet Populations at Opposite Ends of the Age
Distribution}
\shortauthors{Schmidt et al.}

\begin{document}

\title{Resonant and Ultra-short-period Planet Systems are at Opposite
Ends of the Exoplanet Age Distribution}

\author[0000-0001-8510-7365]{Stephen P.\ Schmidt}
\affiliation{William H.\ Miller III Department of Physics \& Astronomy,
Johns Hopkins University, 3400 N Charles Street, Baltimore, MD 21218, USA}

\author[0000-0001-5761-6779]{Kevin C.\ Schlaufman}
\affiliation{William H.\ Miller III Department of Physics \& Astronomy,
Johns Hopkins University, 3400 N Charles Street, Baltimore, MD 21218, USA}

\author[0000-0002-7993-4214]{Jacob H.\ Hamer}
\affiliation{New Jersey State Museum, 205 W State Street, Trenton,
NJ 08608, USA}

\correspondingauthor{Stephen Schmidt}
\email{sschmi42@jh.edu}

\begin{abstract}

\noindent
Exoplanet systems are thought to evolve on secular timescales
over billions of years.  This evolution is impossible to
directly observe on human timescales in most individual systems.
While the availability of accurate and precise age inferences for
individual exoplanet host stars with ages $\tau$ in the interval
$1~\text{Gyr}\lesssim~\tau~\lesssim10~\text{Gyr}$ would constrain this
evolution, accurate and precise age inferences are difficult to obtain
for isolated field dwarfs like the host stars of most exoplanets.
The Galactic velocity dispersion of a thin disk stellar population
monotonically grows with time, and the relationship between age and
velocity dispersion in a given Galactic location can be calibrated
by a stellar population for which accurate and precise age inferences
are possible.  Using a sample of subgiants with precise age inferences,
we calibrate the age--velocity dispersion relation in the Kepler field.
Applying this relation to the Kepler field's planet populations, we find
that Kepler-discovered systems plausibly in second-order mean-motion
resonances have $1~\text{Gyr}\lesssim~\tau~\lesssim2~\text{Gyr}$.
The same is true for systems plausibly in first-order mean-motion
resonances, but only for systems likely affected by tidal dissipation
inside their innermost planets.  These observations suggest that many
planetary systems diffuse away from initially resonant configurations
on secular timescales.  Our calibrated relation also indicates that
ultra-short-period (USP) planet systems have typical ages in the
interval $5~\text{Gyr}\lesssim~\tau~\lesssim6~\text{Gyr}$.  We propose
that USP planets tidally migrated from initial periods in the range
$1~\text{d}\lesssim~P~\lesssim2~\text{d}$ to their observed locations at
$P<1~\text{d}$ over billions of years and trillions of cycles of secular
eccentricity excitation and inside-planet damping.

\end{abstract}


\section{Introduction} \label{sec:intro}

The long-term evolution of exoplanet systems remains a poorly-understood
aspect of exoplanet astrophysics, mostly due to the difficulty of
accurate and precise age inferences for mature main sequence stars
like most exoplanet hosts.  While the occurrence and properties
of planetary systems orbiting young stars enable the study of
short-term planetary system evolution, studies of the long-term
evolution of planetary systems require accurate and precise
ages for mature stars.  Most stellar age inference methodologies
for mature stars like lithium depletion \citep[e.g.,][]{Del90},
moving group membership \citep[e.g.,][]{Bar98,Bar99}, gyrochronology
\citep[e.g.,][]{Barnes2003,2007ApJ...669.1167B}, or stellar activity
\citep[e.g.,][]{Bal95} only work for stars no older than a few Gyr.
New or population-level methods must therefore be used to explore the
evolution of exoplanet systems over billions of years.

We have previously used the Galactic velocity dispersions of thin
disk stellar populations to study the evolution of exoplanet systems
in several contexts.  We showed that short-period hot Jupiters like
those discovered by ground-based transit surveys are destroyed by tides
while their host stars are on the main sequence and that high stellar
obliquity hot Jupiter systems are younger than low stellar obliquity hot
Jupiter systems \citep{HJDestruction,JacobObliquityPaper}.  For smaller
planets, we argued that ultra-short-period (USP) planets are stable to
tidal inspiral and that plausibly resonant systems are relatively young
but older than a few hundred Myr \citep{JacobUSPPaper,JacobMMRPaper}.
While these results have shown that various exoplanet populations are
relatively young or old, to this point we have been unable to obtain
characteristic absolute population ages for these planetary systems.

It is now possible to calibrate the age--velocity dispersion relation
in the volume of the Galaxy searched for transiting planets by the
Kepler space telescope \citep{Borucki2010}, a volume we refer to as
the Kepler field.  \citet{2022Natur.603..599X} published a catalog of
247,104 ages for subgiants amenable to precise isochrone-based age
inference, including 5,078 in the Kepler field.  Gaia Data Release
3 \citep{GaiaDR3Summary,GaiaDR3Validation} provides parallaxes
and proper motions for almost every star in the Kepler field.
Likewise, the Kepler field has been intensively studied by ground-
and space-based spectroscopic surveys, like the California-Kepler
Survey \citep[CKS -][]{2017AJ....154..107P, JohsonCKS2017}, the Large
Sky Area Multi-Object Fiber Spectroscopic Telescope \citep[LAMOST
-][]{2012RAA....12.1197C,2012RAA....12..723Z} Experiment for Galactic
Understanding and Exploration \citep[LEGUE -][]{Den12}, the Apache Point
Observatory Galactic Evolution Experiment \citep[APOGEE -][]{APOGEE},
and Gaia itself \citep{DR3_velocities}.

A calibrated age--velocity dispersion relation in the Kepler
field would be useful for investigating the secular evolution of
small-radius planets in multiple-planet systems.  Before Kepler's
observations, many systems like these were expected to have experienced
convergent Type I migration in their parent protoplanetary disks
\citep[e.g.,][]{TypeIMigration} that would have left them in low-order
mean-motion resonances \citep[e.g.,][]{Ter07}.  However, the period
ratio distribution observed by Kepler does not exhibit this property
\citep{lissauer2011,Fab14}.  Numerous ideas to explain Kepler's
observation of a lack of systems in low-order mean-motion resonance have
been proposed \citep[e.g.,][]{Bar13,Pet13,Cui21,Cha22,Lau22,Liu22}.
\citet{JacobMMRPaper} showed that plausibly second-order mean-motion
resonant systems are younger than the entire population of multiple-planet
systems.  The same is true for plausibly first-order mean-motion resonant
systems, but only if the innermost planet is likely affected by tidal
dissipation.  They showed that these two plausibly resonant populations
are younger than the multiple-planet population but older than a few
hundred Myr as indicated by the activity levels and lithium abundances
of their host stars.  \citet{JacobMMRPaper} were unable to infer absolute
population-level characteristic  mean ages $\tau$\footnote{We use the term
``characteristic'' to indicate that the ages we obtain for our samples
are not calculated as straightforward averages of the individual stellar
ages in our samples, but rather that we use samples' galactic velocity
dispersions as statistical proxies for entire populations represented by
our samples.} using their uncalibrated age--velocity dispersion analyses.
A calibrated age--velocity dispersion relation would enable the inference
of absolute population-level characteristic mean ages for the Kepler
field's planet populations.

A calibrated age--velocity dispersion relation would be even more powerful
for studies of USP planets.  The age of the USP planet population would
inform its formation, because USP planet formation models predict a
range of ages for the USP planet population.  \citet{JacobUSPPaper}
showed that Kepler-discovered USP planets have ages consistent with
field stars.  In accord with this observation, they argued that USP
planets have not recently arrived at their observed locations and
therefore are stable against tidal inspiral.  This interpretation
aligns with ``early arrival'' theories of USP planet migration
\citep[e.g.,][]{Lee17,Pet19,Li2020,Mil20,Bec21,Che22} that advocate for
the arrival of USP planets at their observed locations within a few Gyr.
The \citet{JacobUSPPaper} observation is also consistent with the
possibility that the timescale for USP planets to tidally migrate inward
is more than a few Gyr, making most USP planets recent arrivals.  This
migration regime would be possible under the \citet{2010ApJ...724L..53S}
scenario for USP planet formation in multiple-planet systems, in which
proto-USP planets first inwardly migrate via Type I migration to their
parent protoplanetary disks' magnetospheric truncation radii and then
continue inward via tidal migration after disk dissipation due to cycles
of secular eccentricity excitation and damping inside the planet.

The absolute characteristic mean age of the USP planet
population could differentiate between these possibilities.
If USP planets arrive early as advocated by many authors
\citep[e.g.,][]{Lee17,Pet19,Li2020,Mil20,Bec21,Che22}, then the
characteristic mean age of the USP planet population should be consistent
with all Kepler-discovered small planets.  On the other hand, if USP
planets are recent arrivals, then the characteristic mean age of the
USP planet population should be old.  If USP planets arrive via tidal
migration due to cycles of secular eccentricity excitation and damping
inside the planet, then the population of potential proto-USP planets that
arrived at their parent protoplanetary disks' magnetospheric truncation
radii at orbital periods $1~\text{d} \lesssim P \lesssim 2~\text{d}$
should be systematically younger.

In this article, we execute the tests described above.  Verifying the
\citet{JacobMMRPaper} result, we find that the characteristic mean
ages of plausibly second-order mean-motion resonant systems are older
than few hundred Myr but younger than two Gyr.  The same is true for
plausibly first-order resonant systems, but only if the innermost
planet is likely affected by tidal dissipation.  In contrast, the
population of USP planets has a significantly warmer velocity dispersion
and a characteristic mean age $5 \lesssim \tau \lesssim 6$ Gyr, both
larger than the velocity dispersion and characteristic mean age of the
population of proto-USP planets.  This suggests that systems with a
USP planet are on average older than systems with a proto-USP planet
and supports the idea that USP planets are recent arrivals at their
observed locations.  We describe the construction of our samples in
Section \ref{sec:sample} and our analyses in Section \ref{sec:methods}.
We discuss the implications of our work in Section \ref{sec:disc} and
summarize our conclusions in Section \ref{sec:summ}.

\section{Data} \label{sec:sample}

To study the time evolution of exoplanet systems using the calibrated
age--velocity relation, we identify six types of planetary systems.
From \citet{JacobMMRPaper}, we obtain samples of multiple-planet systems
as well as plausibly first- and second-order mean-motion resonant
systems defined using their $\delta_{\text{res}}$-based criterion.  The
\citet{JacobMMRPaper} $\delta_{\text{res}}$ parameter is a modification
of the mass- and eccentricity-independent $\epsilon$ parameter used by
\citet{Del14b} and \citet{Cha15} to describe proximity to resonance that
normalizes $\epsilon$ by the resonant period ratio to account for the
larger ``widths'' of widely-spaced resonances.  If we were to instead
use the libration width definition in \citet{JacobMMRPaper} for our
first-order resonant sample, we would ultimately find a statistically
indistinguishable result.  We obtain our sample of USP planet systems from
\citet{2014ApJ...787...47S}.  As described in the following paragraph,
we also define a sample of proto-USP planet systems.

It is usually assumed that protoplanetary disks are magnetospherically
truncated at corotation with their host stars.  In that case, the rotation
periods of pre main sequence stars in star-forming regions can be used
to roughly infer the smallest star--planet separations that can result
from Type I migration.  \citet{Rebull2018,Reb20} used continuous K2
data to measure rotation periods for solar-mass pre main sequence stars
with $1 < (V - K_{s})_{0} < 2$ in Taurus and Upper Sco.  They found a
median rotation period $P_{\text{rot}} = 1.2$ d at 3 Myr in Taurus and
$P_{\text{rot}} = 2.1$ d at 8 Myr in Upper Sco.  Since protoplanetary
disks likely dissipate sometime in this age range, these data suggest
that the orbital period corresponding to disk magnetospheric truncation
radii for solar-mass stars during the epoch of planet formation is in the
range $1~\text{d} < P < 2$ d.  We therefore define a sample of planets we
term proto-USP planets as confirmed planets from the Kepler cumulative
planet catalog with planet radii $R_{\text{p}} < 2~R_{\oplus}$ and
orbital periods $1~\text{d} < P < 2$ d that would have place them close
to their parent protoplanetary disk's magnetospheric truncation radii.
We prefer the cumulative KOI list over the DR 25 KOI list due to the
extra level of scrutiny provided by the human vetting that produced the
cumulative catalog.

To calibrate the age--velocity dispersion relation in the Kepler
field, we start with the sample of isochrone-inferred subgiant ages
from \citet{2022Natur.603..599X}.  During the transition from core to
shell hydrogen fusion, small changes in mass correspond to large changes
in easily-observed properties like luminosity or absolute magnitude.
As a result, the subgiant phase of stellar evolution permits the most
precise isochrone-based age inferences.  \citet{2022Natur.603..599X}
fit Yonsei-Yale isochrones \citep{2001ApJS..136..417Y,
2002ApJS..143..499K,2003ApJS..144..259Y,2004ApJS..155..667D} to
\begin{enumerate}
    \item
    photospheric stellar parameters effective temperatures
    $T_{\text{eff}}$, metallicity $[\text{Fe/H}]$, and alpha-to-iron
    ratio $[\alpha/\text{Fe}]$ derived from LAMOST spectroscopy; and
    \item
    Gaia Early Data Release (EDR) 3 parallax-informed absolute
    magnitudes in Gaia EDR3 $G_{\text{BP}}$, $G$, $G_{\text{RP}}$
    and 2 Micron All Sky Survey (2MASS) $JHK_{\text{s}}$ bands
    \citep{GaiaEDR3,Skrutskie2006}.
\end{enumerate} 
To limit the \citet{2022Natur.603..599X} sample to
the Kepler field, we use the sky position of Kepler's
CCDs hosted at STScI's Mikulski Archive for Space Telescopes
(MAST)\footnote{\url{https://archive.stsci.edu/missions/kepler/ffi_footprints/morc_2_ra_dec_4_seasons.txt}}.
This delivers a sample of 5,078 subgiant isochrone-based ages with a
median relative precision of 6\%, corresponding to a typical absolute
age uncertainty of 300 Myr.

To calculate the Galactic velocity dispersion of a sample of stars,
we first calculate each star's $UVW$ velocities.  That calculation
requires as inputs parallax, proper motion, and radial velocity.
We use the parallaxes and proper motions provided by Gaia Data Release
(DR) 3 \footnote{For the details of Gaia DR3 and its data processing,
see \citet{Gaia_mission2016, GaiaEDR3, GaiaDR3Frame, GaiaDR3Summary},
\citet{GaiaEDR3Validation}, \citet{lin21a, lin21b}, \citet{GaiaEDR3XMatch,
GaiaDR3XMatch}, \citet{EDR3photometry} \citet{row21}, \citet{tor21},
and \citet{GaiaDR3Validation}.}.  We find that the Gaia EDR3/DR3 source
identifiers provided by \citet{2022Natur.603..599X} are missing their
last digit.  We correct for this by querying the Gaia archive for the 10
possible corrected \texttt{source\_id}s obtained by appending one digit
to each shortened identifier.  We then select for each subgiant the Gaia
DR3 source that minimizes the on-sky distance between subgiant and Gaia
DR3 source.  We use this corrected list of Gaia DR3 \texttt{source\_id}
information to query the Gaia archive for the necessary astrometric
data.  We impose the constraint \texttt{parallax\_over\_error}
$> 10$ to ensure high-quality astrometry as well as the common
renormalized unit weight error \texttt{ruwe} $< 1.4$ data quality cut
to exclude to the extent possible the influence of unresolved binaries
\citep[e.g.][]{Ziegler2020,lin21a,lin21b}.  This results in a sample of
5,044 Kepler field subgiants that pass our astrometric data quality cuts.

Radial velocities are also necessary to calculate $UVW$ velocities,
and we elect to use Gaia DR3 radial velocities as input to our
velocity dispersion calculation \citep{DR3_velocities}.  Following
\citet{JacobMMRPaper}, we impose the constraints \texttt{rv\_nb\_transits}
$> 10$ and \texttt{rv\_expected\_sig\_to\_noise} $> 5$ to ensure
high-quality radial velocities.  This procedure results in 3,832 subgiants
with Gaia astrometry and radial velocities.  While there are LAMOST
radial velocities available for every star in our sample of subgiants with
precise Gaia astrometry and radial velocities, the typical radial velocity
measurement precision achieved for LAMOST spectra with spectral resolution
$R \approx 1,\!800$ is about 4.3 km s$^{-1}$.  On the other hand, the
typical radial velocity measurement precision achieved for Gaia Radial
Velocity Spectrometer (RVS) spectra with $R \approx 11,\!500$ is about
2.4 km s$^{-1}$.  In addition, we find a zero-point offset between these
LAMOST and Gaia radial velocity measurements of about six km s$^{-1}$.

To determine which of Gaia or LAMOST has the correct zero point,
we use data from APOGEE.  We use data derived from spectra that
were gathered during the third and fourth phases of the Sloan
Digital Sky Survey \citep[SDSS -][]{SDSSIII,SDSSIV} as part of
APOGEE.  These spectra were collected with the APOGEE spectrographs
\citep{Zas13,Zas17,Wil19,Bea21,San21} on the New Mexico State University
1-m Telescope \citep{Hol10} and the Sloan Foundation 2.5-m Telescope
\citep{SDSS2.5m}.  As part of SDSS DR 17 \citep{abd22}, these spectra were
reduced and analyzed with the APOGEE Stellar Parameter and Chemical
Abundance Pipeline \citep[ASPCAP][]{All06,Hol15,Nid15,ASPCAP}
using an $H$-band line list, MARCS model atmospheres,
and model-fitting tools optimized for the APOGEE effort
\citep{Alv98,Gus08,2011ascl.soft09022H,Ple12,Smi13,Smi21,Cun15,She15,2020AJ....160..120J}.
We find no radial velocity zero-point offsets between Gaia and APOGEE or
between Gaia and CKS.  We therefore choose to use Gaia radial velocities
to characterize the the age--velocity dispersion relation in our subgiant
sample.

We follow the same steps described above to calculate the velocity
dispersions of our Kepler-discovered planetary system samples.
We first obtain Gaia DR3 \texttt{source\_id} as well as equatorial
coordinates, parallaxes, and proper motions for our exoplanet
system samples using the 2MASS identifier associated with each
Kepler Input Catalog \citep[KIC;][]{Brown2011} entry and Gaia DR3's
\texttt{tmass\_psc\_xsc\_best\_neighbour} table.  We obtain system radial
velocities for these samples in descending order of priority from the
CKS, APOGEE, and Gaia.  For those planet host stars with Gaia RVS-based
radial velocities, we apply the same data quality cuts described above.
In total, we obtain samples of 60 plausible first-order resonant
systems and 60 plausible second-order resonant systems.  Our sample of
90 USP planet systems has median orbital period $P = 0.68$ d and median
planet radius $R_{\text{p}} = 1.21~R_{\oplus}$, while our sample of 70
proto-USP planet systems has median $P = 1.54$ d and median $R_{\text{p}}
= 1.22~R_{\oplus}$.  For each planet host star in these samples we report
in Table \ref{tab:hosts} the Kepler \& Gaia identifiers, radial velocity
with uncertainty and source, and subsample membership.

\begin{deluxetable*}{lccccccc}
    \centering
    \tabletypesize{\scriptsize}
    \tablewidth{0pt}
    \tablecaption{USP Planet, Proto-USP Planet, and Plausibly/Implausibly
    Resonant System Samples\label{tab:hosts}}
    \tablehead{
    \colhead{Kepler ID} &
    \colhead{Gaia DR3 \texttt{source\_id}} &
    \colhead{RV} &
    \colhead{RV Source} &
    \colhead{Sample}\\
    & & (km s$^{-1}$) &}
    \startdata
    \hline
    1432789 & 2051748659478657152 &$-38.8 \pm 0.1$ & APOGEE & Multiple-planet\\
    1717722 & 2051027792165858304 &$-60.8 \pm 0.1$ & APOGEE & Multiple-planet, USP\\
    1718189 & 2051030231707057024 &$-26.5 \pm 0.1$ & CKS & Second-order MMR\\
    1718958 & 2051019683267537024 &$-24.0 \pm 0.1$ & CKS & Proto-USP\\
    1724719 & 2051721480924900224 &$-15.9 \pm 0.1$ & CKS & Multiple-planet\\
    1871056 & 2051738798233021824 &$-19.2 \pm 0.1$ & CKS & Multiple-planet\\
    1996180 & 2099073667159859840 &$-4.8  \pm 0.1$ & CKS & Multiple-planet\\
    2165002 & 2051832634677771008 &$-39.6 \pm 0.1$ & CKS & Multiple-planet\\
    2299738 & 2052587243251659392 &$-32.9 \pm 4.6$ & Gaia & Proto-USP\\
    2302548 & 2052528625536778624 &$-23.5 \pm 0.1$ & CKS & Multiple-planet\\
    \enddata
    \tablecomments{Tidally-affected first-order resonant systems are
    included in the sample of all first-order resonant systems, and
    the Kepler multiple-planet system sample contains all plausibly
    resonant systems.  This table is sorted by KIC ID and is published
    in its entirety in machine-readable format.}
\end{deluxetable*}

\section{Analysis} \label{sec:methods}

We convert equatorial coordinates, proper motions, parallaxes, and
radial velocities into Galactic space velocities using the \texttt{pyia}
package \citep{adrian_price_whelan_2018_1228136}.  The individual
radial velocity measurement uncertainties are an order of magnitude
smaller than our velocity dispersion measurements, so our analyses
are not limited by radial velocity precision.  We use a Monte Carlo
simulation in which \texttt{pyia} randomly samples from the Gaia
DR3 five-parameter astrometric solution respecting each solution's
covariance.  It also independently randomly samples radial velocities
from a normal distribution with mean and variance as reported in each
radial velocity source.  We sample 100 realizations from each star's
astrometric uncertainty distributions in position, proper motion,
parallax, and radial velocity using \texttt{pyia}.  We then calculate
the velocity dispersion $\sigma$ of each sample
\begin{equation} \label{eq:vd}
    \sigma = \frac{1}{N} \sum \limits_{i=1}^{N} \left[(U_i - \bar{U})^2 + (V_i - \bar{V})^2 + (W_i - \bar{W})^2 \right]^{\frac{1}{2}},
\end{equation}
to construct an ensemble of samples of the entire populations' velocity
dispersions.

\begin{figure*}
    \centering
    \includegraphics[width=\linewidth]{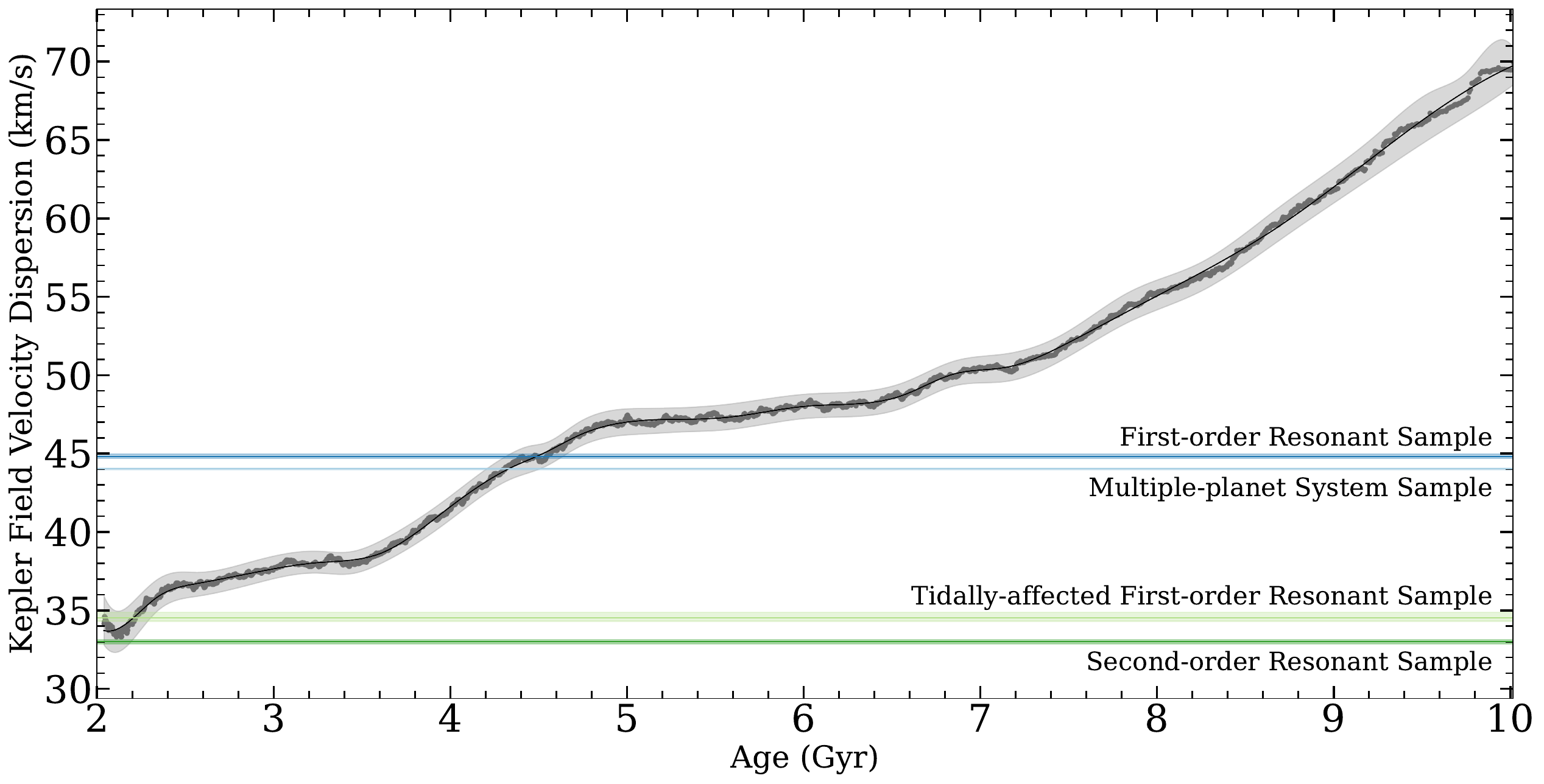}
    \caption{Stellar velocity dispersion as a function of age for the
    Kepler field.  For the sample of subgiants in the Kepler field with
    ages presented in \citet{2022Natur.603..599X}, we first order the
    sample in age.  We then calculate velocity dispersion in consecutive
    windows of 750 stars and plot the result of these calculations as
    overlapping dark gray points.  We plot as the solid black line a
    smoothing spline of these data.  We plot as the gray polygon the
    16th/84th quantile range of the velocity dispersion distribution
    suggested by bootstrap resampling.  We plot as horizontal lines
    the velocity dispersions of the \citet{JacobMMRPaper} samples of
    Kepler-discovered plausibly first- and second-order mean-motion
    resonant systems as well as the subsample of plausibly first-order
    resonant systems with an innermost planet affected by tidal
    dissipation.  For comparison, we plot as the light blue horizontal
    line the velocity dispersion of the complete \citet{JacobMMRPaper}
    sample of Kepler-discovered multiple-planet systems.  The sample
    of second-order resonant systems and the subsample of first-order
    resonant systems with an innermost planet affected by tidal
    dissipation have characteristic mean ages $\tau \lesssim 2$ Gyr.
    On the other hand, the overall sample of first-order resonant systems
    has a similar velocity dispersion and therefore age as the complete
    sample of Kepler-discovered multiple-planet systems.  Combined with
    the \citet{JacobMMRPaper} observation that the characteristic mean
    ages of these resonant populations are older than 500 Myr, the
    implication is that many multiple-planet systems form in mean-motion
    resonances and diffuse away from resonance over about 1 Gyr.}
    \label{fig:mmrcomp}
\end{figure*}

\begin{figure*}
    \centering
    \includegraphics[width=\linewidth]{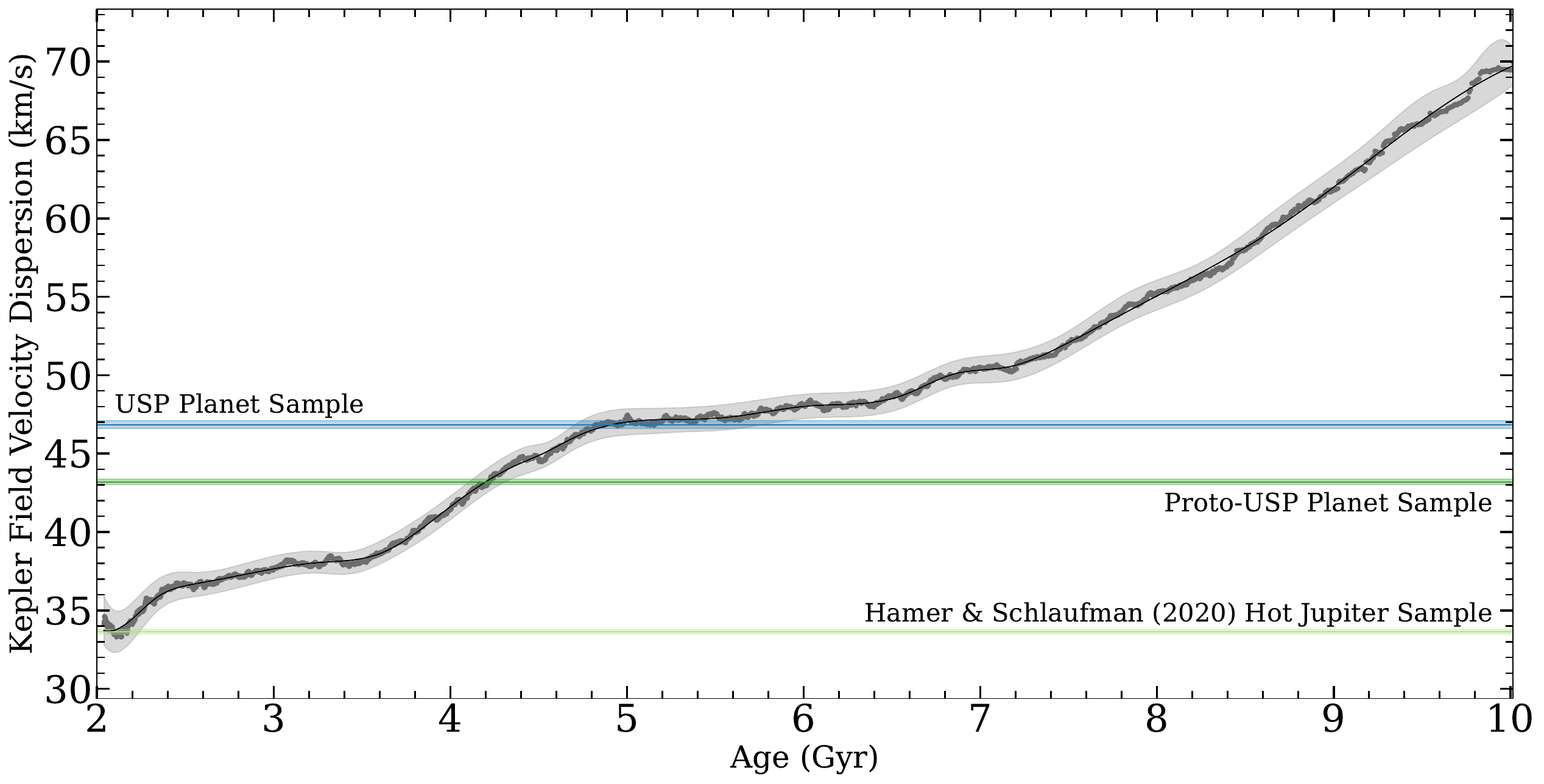}
    \caption{Stellar velocity dispersion as a function of age for
    the Kepler field.  The age--velocity dispersion relation is the
    same as in Figure \ref{fig:mmrcomp}.  We plot as horizontal lines
    the velocity dispersions of the \citet{JacobUSPPaper} sample of
    Kepler-discovered hot Jupiter systems as well as our sample of USP
    planet systems.  We define proto-USP planets as confirmed planets from
    the Kepler cumulative planet catalog with planet radii $R_{\text{p}}
    < 2~R_\oplus$ and orbital periods $1~\text{d} < P < 2$ d, and
    we plot as the dark green horizontal line that sample's velocity
    dispersion.  Our USP planet system sample has characteristic mean age
    $4.7~\text{Gyr} \lesssim \tau \lesssim 5.8$ Gyr, while the proto-USP
    planet sample has characteristic mean age $4.1~\text{Gyr} \lesssim
    \tau \lesssim 4.3$ Gyr.  We argue that the best explanation for this
    age offset is that USP planets tidally migrated from longer-period
    orbits due to cycles of secular eccentricity excitation and tidal
    damping in multiple-planet systems.}
    \label{fig:uspcomp}
\end{figure*}

We use the procedure described above to calculate $UVW$ velocities for our
sample of subgiants.  To generate the age--velocity dispersion relation,
we first sort the subgiant sample by age.  We calculate the mean age
and velocity dispersion in a moving window sample of 750 subgiants,
advance the window by one star, and then repeat until we cover the
entire age range of our subgiant sample.  For each 750-star window,
we obtain 150 bootstrap samples of 750 stars with replacement within
the window and calculate the velocity dispersion of each sample.
We choose a window of 750 to balance time resolution and velocity
dispersion precision.  We report the 16th, 50th, and 84th percentiles
of the resulting velocity dispersion distributions.  In parallel, we
calculate the average age of the 750 stars that make up each window.
To maximize age resolution at the youngest ages, we use smaller windows.
We follow the same procedure described above, but we start each window
with the youngest star but sequentially increase the window width from
150 to 750 as the window advances.  We use univariate smoothing splines
to smooth the curves connecting the 16th, 50th, and 84th percentiles of
the velocity dispersion distributions in each window.  We plot the results
of these calculations in Figures \ref{fig:mmrcomp} and \ref{fig:uspcomp}
and report them in tabular form in Table \ref{tab:ageVD}.

\begin{deluxetable*}{cccc}
    \centering
    \tabletypesize{\scriptsize}
    \tablewidth{0pt}
    \tablecaption{Kepler Field Age--Velocity Dispersion Relation
    \label{tab:ageVD}}
    \tablehead{
    \colhead{Window Average Age} &
    \colhead{Lower Uncertainty} &
    \colhead{Velocity Dispersion} &
    \colhead{Upper Uncertainty} \\
    (Gyr) & (km s$^{-1}$) & (km s$^{-1}$) & (km s$^{-1}$)}
    \startdata
    2.04007 &  1.35312 &  34.17309	 &  1.58540\\
    2.04181 &  1.21826 &  34.28341	 &  1.22931\\
    2.04354 &  1.29945 &  34.62238	 &  1.51263\\
    2.04525 &  1.50722 &  34.53612	 &  1.30693\\
    2.04695 &  1.58769 &  34.56045	 &  1.53181\\
    2.04868 &  1.52845 &  34.46360	 &  1.48397\\
    2.05040 &  1.20451 &  34.35825	 &  1.22901\\
    2.05210 &  1.54166 &  34.30429	 &  1.42783\\
    2.05380 &  1.49620 &  34.34044	 &  1.71104\\
    2.05549 &  1.37817 &  34.26295	 &  1.35232\\
    \enddata
    \tablecomments{This table is ordered by average age in ascending
    order and is published in its entirety in machine-readable format.
    Upper uncertainty refers to the difference between the 84th and
    50th percentiles, and the lower uncertainty refers to the difference
    between the 50th and 16th percentiles.}
\end{deluxetable*}

We use the age--velocity dispersion relation derived above to calculate
velocity dispersion-based characteristic mean ages for the six planet
populations represented in Table \ref{tab:hosts}.  In particular,
we infer lower and upper limits for these characteristic mean ages
by identifying the range over which our inferred planet population
mean velocity dispersions overlap with the 1-$\sigma$ range of the
subgiant-based Kepler field age--velocity dispersion relation.  We do
the same for the Kepler field hot Jupiter and USP planet system samples
from \citet{JacobUSPPaper}.  We give the resulting characteristic mean
age intervals in Table \ref{tab:jacobresults}.

\begin{deluxetable*}{lcccc}
\centering
\tabletypesize{\scriptsize}
\tablewidth{0pt}
\tablecaption{Velocity Dispersions for Exoplanet Populations in the
Kepler Field \label{tab:jacobresults}}
\tablehead{\colhead{Sample} & \colhead{Number of Systems} &
\colhead{Velocity Dispersion} &
\colhead{Age Range} &
\colhead{Source} \\
& & (km s$^{-1}$) & (Gyr) &}
\startdata
\hline
Ultra-short-period Planet Systems & 68 & $47.0^{+0.1}_{-0.1}$ & $(4.7, 5.9)$ &\citet{JacobUSPPaper}\\
Hot Jupiter Systems & 24 & $33.6^{+0.2}_{-0.2}$ & $< 2.2$ & \citet{JacobUSPPaper} \\
Ultra-short-period Planet Systems &  90 & $46.8^{+0.3}_{-0.2}$ & $(4.7, 5.8)$ & This work \\
Proto-ultra-short-period Planet Systems & 70 & $43.2^{+0.2}_{-0.2}$ & $(4.1, 4.3)$ & This work \\
Multiple-planet Systems & 563 & $44.0^{+0.1}_{-0.1}$ & $(4.2, 4.5)$ & This work \\
Plausibly First-order Resonant Systems & 60 & $44.8^{+0.2}_{-0.1}$ & $(4.3, 4.6)$ & This work \\
Plausibly Second-order Resonant Systems & 60 & $33.0^{+0.1}_{-0.1}$ & $(0.5, 2.2)$ & This work\\
Tidally-affected Plausibly First-order Resonant Systems & 9 & $34.5^{+0.3}_{-0.2}$ & $(0.5, 2.3)$ & This work\\
\enddata
\end{deluxetable*}

\section{Discussion} \label{sec:disc}

We find that plausibly second-order mean-motion resonant and plausibly
first-order mean-motion resonant systems likely affected by tidal
dissipation discovered by Kepler have characteristic mean ages in the
range $0.5~\text{Gyr} \lesssim \tau \lesssim 2$ Gyr.  We also find that
USP planet systems discovered by Kepler have characteristic mean ages
in the range $4.7~\text{Gyr} \lesssim \tau \lesssim 5.8$ Gyr, a value
offset to older ages than we found for our sample of proto-USP systems
$4.1~\text{Gyr} \lesssim \tau \lesssim 4.3$ Gyr.  We emphasize that this
is a characteristic mean age, not that there are no young USP planet
systems.  Indeed, the first USP planet discovered CoRoT-7 b appears to
be in a relatively young system \citep{CoRoT-7I,CoRoT-7II}.

The young characteristic mean ages of plausibly second-order mean-motion
resonant and plausibly first-order mean-motion resonant systems likely
affected by tidal dissipation supports the hypothesis put forward
by \citet{JacobMMRPaper} that systems of small-radius planets often
form resonant and diffuse away from resonances on secular timescales.
The absolute ages we infer here provide further evidence against
short-timescale explanations for the apparent lack of planet pairs
close to low-order mean-motion resonance in the Kepler-discovered period
distribution.  These short-timescale processes include in situ formation
\citep[e.g.,][]{Pet13}, interactions between planets and protoplanetary
disk density waves caused by outer planets \citep[e.g.,][]{Bar13,Cui21},
planet--planet interactions mediated by protoplanetary disk-driven
eccentricity damping \citep[e.g.,][]{Cha22,Lau22}, and progressive
protoplanetary disk dissipation-driven disruptions \citep[e.g.,][]{Liu22}.
Our inference provides for the first time the characteristic 1 Gyr
timescale for the diffusion of initially resonant systems away from
resonances.

Our conclusion that the population of USP planets is older than the
population we define as proto-USP planets is inconsistent with USP
planet formation models in which USP planets arrive at their observed
locations in less than about 1 Gyr.  We rule out USP planet formation
mechanisms that take place while protoplanetary disks are present,
like the scenario advocated by \citet{Bec21} as a consequence of
protostellar outburst-driven accretion events.  Our observations are
likewise inconsistent with USP planet formation as a consequence of
the gravitational influence of an oblate host star's $J_2$ quadrupolar
potential \citep{Li2020,Che22}, as that scenario requires USP planets to
have arrived at their observed locations in less than 1 Gyr.  Obliquity
tide-driven migration is also expected to have a timescale shorter than
1 Gyr \citep{Mil20}.

Though eccentricity excitation-driven migration has been a frequent
explanation for USP planet formation, different models predict a range
of migration timescales.  In general, a small number of large-amplitude
eccentricity excitations will lead to short migration timescales
\citep[e.g.,][]{Pet19}.  These short migration timescales are disfavored
by our conclusion that USP planets take many Gyr to arrive at their
observed locations.  On the other hand, frequent cycles of low-amplitude
eccentricity excitation will lead to long migration timescales that are
consistent with our observations \citep[e.g.,][]{2010ApJ...724L..53S}.
Tidal migration resulting from the dissipation of orbital energy in
an especially dissipative host star would also be consistent with our
observations \citep[e.g.,][]{Lee17}.

\begin{figure*}
    \centering
    \includegraphics[width=\linewidth]{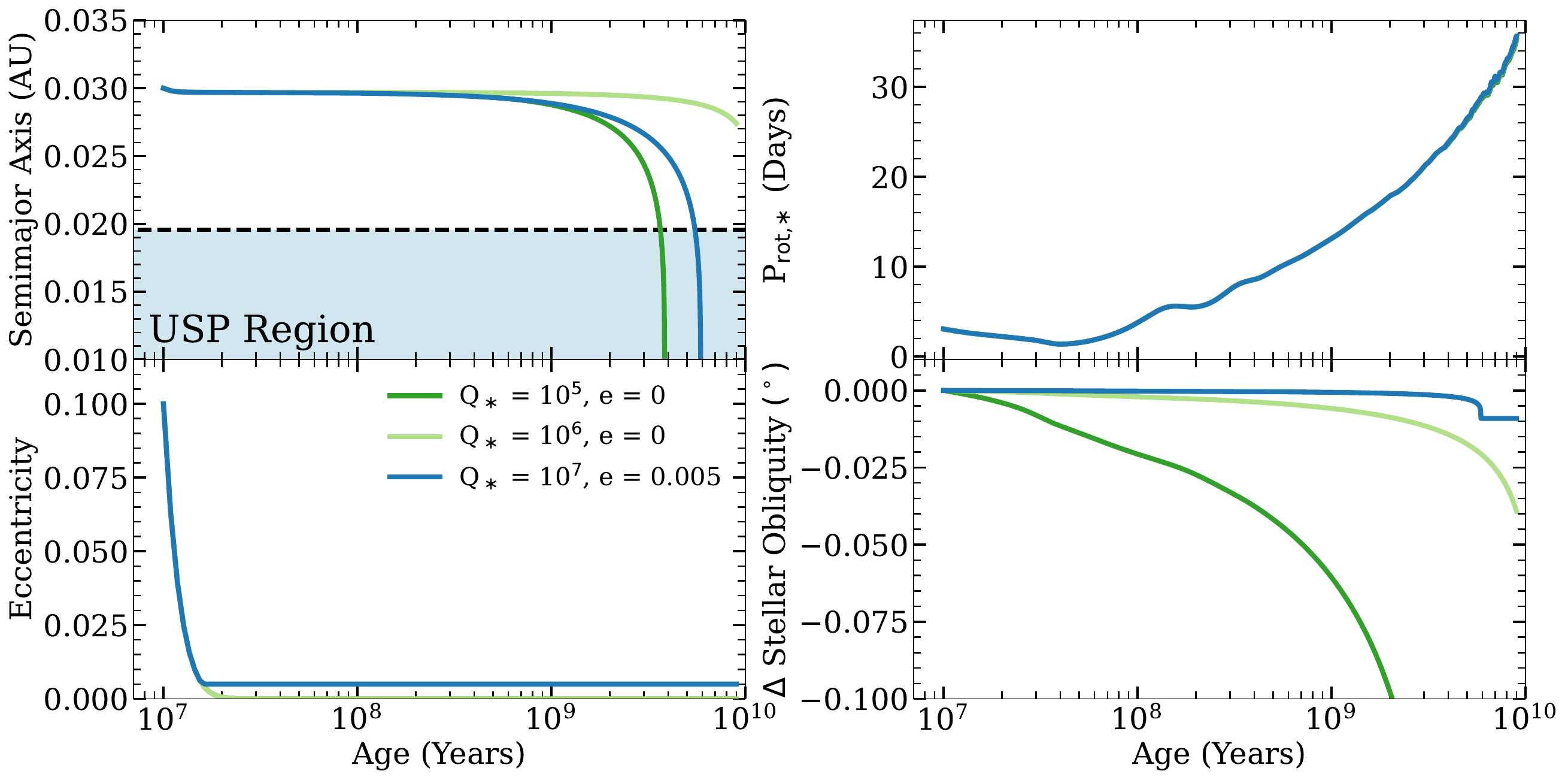}
    \caption{Model for the self-consistent tidal evolution of a typical
    USP planet system.  We plot as colored lines the semimajor axis
    (top left), eccentricity (bottom left), stellar rotation period
    (top right), and difference in stellar obliquity from the initial
    value of $20^\circ$ (bottom right) as functions of time in years.
    Our model uses the system of ODEs presented in \citet{Lec10} that
    solves the coupled tidal evolution of the rotation and revolution
    of both a planet and its host star.  We account for the rotational
    evolution of the host star due to stellar evolution and winds
    using a solar-composition, median-rotation $M_{\ast} = 1~M_{\odot}$
    stellar model from the rotational stellar evolution grid presented
    in \citet{Ama19}.  We model the median USP planet with $R_{\text{p}}
    = 1.2~R_{\oplus}$, $2.0~M_{\oplus}$, and $Q_{\text{p}}' = 10^2$.  We
    plot in blue the solution with an assumed stellar tidal quality factor
    $Q_{\ast}' = 10^7$ as appropriate for solar-type main sequence stars
    \citep{Bar20} and a nonzero but unobservable excited eccentricity $e
    = 0.005$.  Given these parameters, the rotation and revolution of the
    planet are quickly synchronized and the planetary obliquity is quickly
    damped \citep[e.g.,][]{Lec10}.  To evaluate whether tidal dissipation
    in the star could reproduce our observational result we plot as dark
    and light green lines models calculated using the same ODE system,
    but with no eccentricity excitation and smaller $Q_{\ast}' = 10^6$
    (light green) and $Q_{\ast}' = 10^5$ (dark green).  This model
    corroborates our observational evidence that it takes billions of
    years and trillions of orbits for a rocky planet to tidally migrate
    and become a USP planet as a consequence of tidal dissipation inside
    the planet.}
    \label{fig:odes}
\end{figure*}

To investigate the plausibility of the scenario that proto-USP planets
tidally migrate to become USP planets due to cycles of eccentricity
excitation and dissipation, we integrate the coupled system of ordinary
differential equations (ODEs) modeling the tidal evolution process
presented in \citet{Lec10}.  This system of ODEs is based on the complete
tidal evolution equations of the \citet{Hut81} model and valid at any
order in eccentricity, obliquity, and spin.  For the host star of our
model system, we use the self-consistent rotational stellar evolution
model for a $M_{\ast} = 1~M_{\odot}$, solar-composition, median-rotating
star presented in \citet{Ama19}.  The median USP planet in our sample
has planet radius $R_{\text{p}} = 1.2~R_{\oplus}$, corresponding to a
planet mass $M_{\text{p}} = 2.0~M_{\oplus}$ assuming the \citet{Zen19}
mass--radius relation for an Earth-like composition.  We assume the
system begins its post protoplanetary disk dissipation evolution with
eccentricity $e = 0.1$ at $a = 0.030$ AU, corresponding to $P = 1.9$ d.
We further assume that the planet has specific tidal quality factor
$Q_{\text{p}}' = 10^2$, its host star has a stellar tidal quality factor
$Q_{\ast}' = 10^{7}$, and the system's eccentricity cannot fall below
a non-zero but still unobservably small value $e = 0.005$.  Under these
conditions, the planet reaches a period $P < 1$ d in 5 Gyr, a timescale
consistent with our observational result.  We plot the result of this
calculation in Figure \ref{fig:odes}.  We expect Ohmic dissipation to have
a negligible effect on the orbital evolution of the planet, as the amount
of orbital energy dissipated in this way is much, much less than the
amount of orbital energy dissipated through tides \citep[e.g.,][]{Wu13}.

As $Q_{\text{p}}'$ is sensitive to interior structure and tidal forcing
forcing frequency \citep{Tob19}, it may vary from our assumed value
$Q_{\text{p}}' = 10^2$ within the range $80 < Q_{\text{p}}' < 280$.
This interval is bounded below by measurements of $Q_{\text{p}}'$ for Mars
\citep{Lai07, Lai21} and above by predictions for Earth's $Q_{\text{p}}'$
if it lacked oceans \citep{Ray01}.  This range of possible values could
lead to migration timescales that differ by up to a factor of two (i.e.,
a few Gyr), as $da/dt$ from \citet{Lec10} is linear in $Q_{\text{p}}'$.
Likewise, changes of 0.001 to the system's minimum eccentricity would
lead to a similar lengthening or shortening of its migration timescale.
We emphasize that in addition to a range of possible $Q_{\text{p}}'$,
the excited eccentricity is not a precisely known quantity but rather
an empirical choice to match our observation.

We also argue that the minimum eccentricity we assumed in our model
as a consequence of secular eccentricity oscillations in multiple
planet systems is also plausible.  We use the \texttt{celmech} code
\citep{celmech} to study the secular behavior of USP planet systems
resembling the USP systems with additional, longer-period transiting
planets identified by \citet{2014ApJ...787...47S}.\footnote{These
systems are Kepler-10, Kepler-487, Kepler-607, Kepler-990, Kepler-1315,
Kepler-1322, Kepler-1813, Kepler-1814, Kepler-1834, and Kepler-1977.} In
the absence of tidal dissipation, the secular eccentricity excitation
caused by planets exterior to a proto-USP planet are predicted to
easily be large enough to sustain the excited eccentricity we assume in
our calculations.  The self-consistent calculation of the eccentricity
excitation and tidal dissipation in an $N$-body model of a proto-USP
planet system over the necessary timescales is computational expensive
and much more complicated.  We suggest that such a calculation should
be the subject of future investigation.

Our preferred model relies on a sustained, very small, but non-zero
eccentricity that will be very hard to measure for a USP or proto-USP
planet system.  While Doppler measurements lack the necessary precision,
the precise timing of transits and secondary eclipses could in principle
result in a more attainable observational eccentricity constraint for a
USP planet system.  Full orbits of several USP planets orbiting solar-type
host stars have been or are currently scheduled for observation by the
James Webb Space Telescope (JWST) in Cycles 1 or 2, including K2-22
b \citep{San15}, K2-141 b \citep{Mal18}, and TOI-561 b \citep{Lac21}.
To evaluate the eccentricity constraints that may be possible with these
observations, we use Equation (6) from \citet{Wal50} for the tangential
component of the eccentricity
\begin{equation}
    e \cos{\omega} = \frac{\pi}{P} \left(\frac{t_2 - t_1 - P/2}{1 + \csc^2{i}}\right),
\end{equation}
where $e$, $\omega$, $P$, and $i$ are the system's eccentricity,
argument of periastron, orbital period, and inclination.  The quantity
$t_2 - t_1$ is the time difference between the transit and eclipse
midpoints.  In the ideal situation where $e$ is entirely tangential
(i.e., $\cos{\omega} = 1$), robust measurements would require
eccentricity inference precisions of about 0.001.  In systems with $i <
89.9^\circ$, eccentricity uncertainties are always larger than inclination
uncertainties so inclination uncertainties will be the limiting factors
for eccentricity inferences.  To this point, inclination uncertainties
for the best-characterized USP planet systems have always been larger than
$0.1^\circ$ \citep[e.g.,][]{Fog14, Sta17, Van17, Bou18, Bri23}.  This is
also the case for the larger sample of \citet{2014ApJ...787...47S} Kepler
discoveries that has a median inclination uncertainty of $3.5^\circ$.
The net result is that the sustained eccentricity necessary for our
tidal migration scenario is at present too small to be observed.

Our conclusion that the characteristic mean age $4.7~\text{Gyr} \lesssim
\tau \lesssim 5.8$ Gyr of the population of Kepler-discovered USP planets
can also in principle be explained by the USP planet formation scenario
invoking tidal dissipation inside host stars advocated by \citet{Lee17}
if protoplanetary disks are truncated at $P \lesssim 1$ d.  The analyses
presented in that article relied on hypothesized disk magnetospheric
truncation radii inferred from ground-based stellar rotation periods
for young stars in the Orion Nebular Cluster, NGC 2362, and NGC 2547.
Those rotation periods were derived from ground-based observations
that, while sensitive to the large-amplitude variations typical
of young low-mass star light curves, were likely insensitive to the
small-amplitude variations typical of young solar-mass star light curves.
Rotation periods based on ground-based data are also susceptible to
aliasing with the one-day observing cadence of all single-location
ground-based observations.  The sample of rotation periods for stars in
the mass range $0.5~M_{\odot} < M_{\ast} < 1.4~M_{\odot}$ shown in Figure
2 of \citet{Lee17} are almost certainly biased toward the low-mass end of
that mass range because lower-mass stars are both more numerous and have
larger-amplitude variations in their light curves more easily detected
in ground-based data.  They may also be aliased to $P_{\text{rot}} = 1$
d because of observing cadence.

As we argued in Section \ref{sec:sample}, continuous space-based K2
observations of Taurus and Upper Sco analyzed in \citet{Rebull2018,Reb20}
suggest that disks around solar-mass protostars are truncated somewhere
in the range $1~\text{d} \lesssim P \lesssim 2$ d.  To evaluate the
\citet{Lee17} scenario in light of these new data, we again solve
the \citet{Lec10} system of ODEs assuming the same stellar and planet
properties but with two changes: we allow the system's eccentricity to go
to zero and vary the stellar specific tidal quality factor $Q_{\ast}'$.
We find that in this case a proto-USP planet can evolve into a USP
planet in 5 Gyr only when its host star has $Q_{\ast}' \sim 10^{5}$,
about an order of magnitude smaller (i.e., more dissipative) than
predicted by theoretical models in this planet mass, stellar mass,
and orbital period range \citep[e.g.,][]{Ogi07, Bar20, Wei24}.  We plot
this calculation for both $Q_{\ast}' = 10^{5}$ and $Q_{\ast}' = 10^{6}$
in Figure \ref{fig:odes}.  Unless stars are significantly more dissipative
than expected with a typical $Q_{\ast}' \approx 2 \times 10^{5}$, then the
scenario in which proto-USP planets start at their parent protoplanetary
disk's magnetospheric truncation radii and migrate inward due to tidal
dissipation inside their host star is inconsistent with our observations.

The stellar tidal quality factor required for tidal dissipation inside
a proto-USP planet's host star to explain our observation is similar
to the measured value $Q_{\ast}' \approx 2 \times 10^5$ for WASP-12
\citep{Mac16, Pat17, Yee20, Tur21}, the only plausibly main sequence
star with a directly inferred $Q_{\ast}'$ value.\footnote{Kepler-1658
b has also been shown to be experiencing orbital decay \citep{Vis22},
but its host star is unambiguously a subgiant.}  However, WASP-12's
current $Q_{\ast}'$ lies at the lower end of the range of $Q_{\ast}'$
inferred for hot Jupiter host stars \citep[e.g.,][]{Jac08, Lan11,
Hus12, Ada24}.  In addition, a hot Jupiter system's orbital period, host
star structure, and planet mass have been theoretically shown to affect
$Q_{\ast}'$ \citep[e.g.,][]{Wei12, Wei24}.  These models suggest that
the $Q_{\ast}'$ for WASP-12 is at least an order of magnitude smaller
(i.e., more dissipative) than for the host stars of our USP and proto-USP
planet samples.  It is also plausible that WASP-12 is a subgiant and
consequently more dissipative than the main sequence host stars in our
samples \citep{Wei17, Bai19}.  We therefore argue that the value of
$Q_{\ast}'$ inferred for the WASP-12 system it is not representative of
the host stars in our USP and proto-USP planet samples.

\section{Conclusion} \label{sec:summ}

We find that the population of Kepler-discovered multiple-planet systems
has a characteristic mean age $4.2~\text{Gyr} \lesssim \tau \lesssim
4.5$ Gyr.  On the other hand, the population of Kepler-discovered
plausibly second-order mean-motion resonant planetary systems has a
characteristic mean age in the range $0.5~\text{Gyr} \lesssim \tau
\lesssim 2$ Gyr.  The same is true for Kepler-discovered plausibly
first-order mean-motion resonant planetary systems, but only for systems
likely affected by tidal dissipation inside their innermost planets.
We conclude that many planetary systems form in resonance and then
diffuse away from resonance with a characteristic timescale of 1 Gyr.
We show that the population of Kepler field USP planet systems has a
characteristic mean age $4.7~\text{Gyr} \lesssim \tau \lesssim 5.8$
Gyr that is older than the characteristic mean age of the population
of proto-USP planets $4.1~\text{Gyr} \lesssim \tau \lesssim 4.3$ Gyr.
The older age of USP planets suggests that they have only recently
arrived at their observed locations, an observation that is inconsistent
with models of USP planet formation in which USP planets arrive at $P
\lesssim 1$ d in less than 1 Gyr.  Among the range of models proposed for
USP planet formation, we suggest that the formation of USP planets via
tidal migration from initial periods in the range $1~\text{d} \lesssim
P \lesssim 2~\text{d}$ to their observed locations at $P < 1~\text{d}$
as a consequence of billions of years and trillions of cycles of secular
eccentricity excitation and tidal damping inside the tidally migrating
planet is most consistent with the available data.

\section*{Acknowledgments}

We thank the anonymous referee for their helpful and insightful comments.
We thank Colin Norman, Sam Grunblatt, Daniel Thorngren, and David
Sing for helpful insight about our findings.  This material is based
upon work supported by the National Science Foundation under grant
number 2009415.  This work has made use of data from the European Space
Agency (ESA) mission {\it Gaia} (\url{https://www.cosmos.esa.int/gaia}),
processed by the {\it Gaia} Data Processing and Analysis Consortium (DPAC,
\url{https://www.cosmos.esa.int/web/gaia/dpac/consortium}).  Funding for
the DPAC has been provided by national institutions, in particular the
institutions participating in the {\it Gaia} Multilateral Agreement.  This
research has made use of the SIMBAD database, operated at CDS, Strasbourg,
France \citet{Wenger2000}.  This research has made use of the VizieR
catalog access tool, CDS, Strasbourg, France.  The original description
of the VizieR service was published in \citet{VizieR}.  This research
has made use of the NASA Exoplanet Archive \citep{ExoplanetArchive},
which is operated by the California Institute of Technology, under
contract with the National Aeronautics and Space Administration under the
Exoplanet Exploration Program.  Funding for SDSS-III has been provided
by the Alfred P. Sloan Foundation, the Participating Institutions, the
National Science Foundation, and the U.S. Department of Energy Office
of Science.  The SDSS-III web site is \url{http://www.sdss3.org/}.
SDSS-III is managed by the Astrophysical Research Consortium for the
Participating Institutions of the SDSS-III Collaboration including the
University of Arizona, the Brazilian Participation Group, Brookhaven
National Laboratory, Carnegie Mellon University, University of Florida,
the French Participation Group, the German Participation Group, Harvard
University, the Instituto de Astrofisica de Canarias, the Michigan
State/Notre Dame/JINA Participation Group, Johns Hopkins University,
Lawrence Berkeley National Laboratory, Max Planck Institute for
Astrophysics, Max Planck Institute for Extraterrestrial Physics, New
Mexico State University, New York University, Ohio State University,
Pennsylvania State University, University of Portsmouth, Princeton
University, the Spanish Participation Group, University of Tokyo,
University of Utah, Vanderbilt University, University of Virginia,
University of Washington, and Yale University.  Funding for the
Sloan Digital Sky Survey IV has been provided by the Alfred P. Sloan
Foundation, the U.S.  Department of Energy Office of Science, and the
Participating Institutions.  SDSS-IV acknowledges support and resources
from the Center for High Performance Computing at the University of
Utah.  The SDSS website is www.sdss4.org.  SDSS-IV is managed by the
Astrophysical Research Consortium for the Participating Institutions
of the SDSS Collaboration including the Brazilian Participation Group,
the Carnegie Institution for Science, Carnegie Mellon University, Center
for Astrophysics | Harvard \& Smithsonian, the Chilean Participation
Group, the French Participation Group, Instituto de Astrof\'isica
de Canarias, The Johns Hopkins University, Kavli Institute for the
Physics and Mathematics of the Universe (IPMU) / University of Tokyo,
the Korean Participation Group, Lawrence Berkeley National Laboratory,
Leibniz Institut f\"ur Astrophysik Potsdam (AIP),  Max-Planck-Institut
f\"ur Astronomie (MPIA Heidelberg), Max-Planck-Institut f\"ur Astrophysik
(MPA Garching), Max-Planck-Institut f\"ur Extraterrestrische Physik (MPE),
National Astronomical Observatories of China, New Mexico State University,
New York University, University of Notre Dame, Observat\'ario Nacional /
MCTI, The Ohio State University, Pennsylvania State University, Shanghai
Astronomical Observatory, United Kingdom Participation Group, Universidad
Nacional Aut\'onoma de M\'exico, University of Arizona, University
of Colorado Boulder, University of Oxford, University of Portsmouth,
University of Utah, University of Virginia, University of Washington,
University of Wisconsin, Vanderbilt University, and Yale University.
This paper includes data collected by the Kepler mission and obtained from
the MAST data archive at the Space Telescope Science Institute (STScI).
Funding for the Kepler mission is provided by the NASA Science Mission
Directorate.  STScI is operated by the Association of Universities
for Research in Astronomy, Inc., under NASA contract NAS 5–26555.
This research has made use of NASA's Astrophysics Data System.

\vspace{5mm}
\facilities{ADS, CDS, Exoplanet Archive, Gaia, Kepler, MAST, Sloan}

\software{astropy \citep{AstropyI, AstropyII, AstropyIII},
\texttt{celmech} \citep{celmech},
\texttt{matplotlib} \citep{hunter2007matplotlib},
\texttt{numpy} \citep{harris2020array},
\texttt{pandas} \citep{McKinney_2010, reback2020pandas},
\texttt{pyia} \citep{adrian_price_whelan_2018_1228136},
\texttt{scipy} \citep{jones2001scipy,2020SciPy-NMeth},
\texttt{TOPCAT} \citep{TOPCAT}}

\bibliography{article_bibliography}{}
\bibliographystyle{aasjournal}

\end{document}